# Design for Online Deliberative Processes and Technologies: Towards a Multidisciplinary Research Agenda

**Lu Xiao**
University of West Ontario
North Campus Building 256
London, Ontario, Canada
lxiao24@uwo.ca

**Weiyu Zhang**
National University of Singapore
11 Computing Drive, 03-24
Singapore, 117416
weiyu.zhang@nus.edu.sg

**Anna Przybylska**
University of Warsaw
Instytut Socjologii UW, 00-927
Warszawa, ul. Karowa 18
a.przybylska@uw.edu.pl

**Anna De Liddo**
Open University
Walton Hall, MK7 6AA Milton
Keynes, UK a.deliddo@open.ac.uk

**Gregorio Convertino**
Informatica Corporation
2100 Seaport Boulevard, Redwood
City, CA, USA 94063
gconvertino@gmail.com

**Todd Davies**
Stanford University
Stanford, CA 94305-2150
davies@stanford.edu

**Mark Klein**
MIT
5 Cambridge Center NE25-754
Cambridge MA 02139
m_klein@mit.edu

## Abstract
There has been rapidly growing interest in studying and designing online deliberative processes and technologies. This SIG aims at providing a venue for continuous and constructive dialogue between social, political and cognitive sciences as well as computer science, HCI, and CSCW. Through an online community and a modified version of world cafe discussions, we contribute to the definition of the theoretical building blocks, the identification of a research agenda for the CHI community, and the network of individuals from academia, industry, and the public sector who share interests in different aspects of online deliberation.

## Author Keywords
Argumentation; collective intelligence; deliberation system; ideation; social innovation

## ACM Classification Keywords
H.5.3. Computer-supported cooperative work

## Introduction
Online deliberation is an important and increasingly common Internet phenomenon that has drawn attention from researchers in various disciplines. A series of events, projects and workshops have been convened by organizers of this SIG. From 2003-2010 there were four International Conferences on Online

Deliberation [1]. Over the past five years, gatherings have moved to workshops hosted by conferences in specialized - although interdisciplinary - areas such as Computer Supported Cooperative Work (CSCW) and Human-Computer Interaction (HCI) (e.g., ACM CSCW 2012 [2], COOP 2012, Communities and Technologies 2013 [3]); Social, Political and Media Communication [4]; Collective Intelligence, Sensemaking, and Computer-Supported Argumentation [5, 6]; and Natural Language Processing [7]). Examples of recent studies of large-scale deliberation emerged from these workshops can be found in a 2015 journal issue [8].

The trend toward fragmentation across specialized areas partly reflects the difficulties encountered in the early attempts to bring researchers from many disciplines together. The fundamental assumptions (such as the meaning of "deliberation") differ particularly across the computer science and social science communities.  Nonetheless, the problem of designing deliberative processes and technologies remains inherently interdisciplinary. The goal of this SIG meeting is to map these connections and set the research landscape for the HCI community to study and design online deliberative processes and technologies.

**Format**
Our SIG meeting will be held in the form of conducting an interactive panel session both online and offline. First, we introduce the panel's aim and plan of work, as well as an overview of the existing efforts from both social sciences and HCI fields (20 minutes). Then, we will have moderated group discussions including both roundtable discussions at the CHI venue (approximately five tables) and one online discussion group (40 minutes). Thirdly, the moderator of each group will summarize the discussion and the organizers will conclude the meeting with a general wrap-up (20 minutes). One outcome of the SIG meeting will be a report based on registered and transcribed discussions.

Prior to the conference: Between March 2 and April 10, we will set up a website, connected with the CHI conference website, to engage researchers and practitioners in agenda building for the panel session, especially the discussion topics. On the website we will explain the goal of this SIG meeting and its format. We will also provide several discussion topics and a couple of questions that are associated with each topic. We will ask our website's visitors while registering:

- To choose between the online and offline (CHI's venue) form of participation in the panel,
- To add topics and questions for the group discussions, if they wish,
- To affirm or add to questions proposed by the organizers,
- To comment or edit text introducing topics.

Based on the visitors' input, we will finalize the topics and questions to be discussed at the meeting (either through roundtables at the conference or online through our deliberation web site). We are aiming at five discussion topics and no more than three questions for each topic. The selected topics and questions will be posted on the website prior to the meeting.

On the day of the SIG meeting: We will have online and offline discussions in parallel. Each group will select a moderator who plays the role of a coordinator and someone who will report the discussion back to the meeting attendees. We will use virtual channels to live broadcast our roundtable discussions.  Virtual broadcasting channels will be selected based on several criteria, such as accessibility, and the organizers' prior experiences with the channels.

**Discussion Topics at the SIG Meeting**

As described above, the discussion topics and questions will be decided in a cooperative fashion by the

organizers and participants. The list of topics that we will post on the website as starting points include:
- Applications of existing technologies, such as email, blogs, wikis, chat, web forums, Q&A sites, peer-rated news aggregators, social network sites, ideation and argumentation tools, for deliberation;
- Criteria and methodology to evaluate the quality of online deliberation;
- Access to information content and formats, and their impact on deliberation outcomes;
- Social and psychological motivations of users for contributing to online deliberation, and technologies such as gamification to motivate participation;
- Deliberation processes, such as internal reflection, group dynamics, moderation, and technological functions, and their impact on deliberation outcomes;
- Collective and individual level deliberation outcomes, including knowledge acquisition, opinion change, affective responses, and intent to collaborate.

## Attendees
Our tentative hosting website is debatehub.net. We propose that the website be connected with the website of the conference. We will disseminate information about the SIG panel and its website through email lists, websites of organizations like CHI, ECREA, AoIR, ICA, APSA, networks like Digital Democracy Network DEL, and email lists for our organizations and institutions.

We will also recruit attendees by emailing all authors of papers presented at previous workshops of "Large-scale idea management and deliberation systems", "Collective Intelligence for the Common Good", "Online Deliberation Emerging Tools", "Argumentation Mining", "Deliberation: values, processes, institutions", "New perspectives for dialogue: The model of deliberation and ICT tools in decision-making processes", four rounds of online deliberation conferences, and interested members of the International Communication Association and American Political Science Association. Additionally, we will advertise the SIG via social media channels such as Facebook and Twitter.

## Expected outcome
We anticipate that this SIG meeting will help bring together a diversity of researchers and practitioners and lay common ground for the definition and organization of research and design for online deliberative processes and technologies. Specifically, we expect to
- provide more rigorous definitions of the topics under scrutiny in relation to existing theories, methods, and technologies (e.g., to set criteria for "good" online deliberation to guide design);
- map the space of design problems and promising solutions relevant to researchers and practitioners developing online deliberation technologies; and
- identify horizontal challenges, and the strategies to approach them, when studying and designing online deliberation (e.g. licenses, formalizing, gamification, visualization).

After the SIG meeting, we will continue to build a multidisciplinary network for studying and designing technology-mediated deliberation (e.g., build an email list, open a Facebook group, dedicate a website to the community, organize regular workshops, publish special issues and books, promote collaborations on grants and projects, and generate patents and commercial products). We anticipate that the fostered communication and collaboration among researchers will promote more awareness of research and practice from different domains, leading to a more comprehensive understanding of the design and evaluation of deliberation technologies.

A report with the conclusions from the SIG meeting will be made available through its website, the organizers' websites, and potentially through a publication.

**Organizers of the SIG meeting**
The seven organizers truly represent the multidisciplinary and international nature of the SIG. We are from the US, Canada, Poland, Singapore, and the UK. Four of us are from the HCI community, and three from the communication discipline. Six of us are from academia and one is from industry. Four were co-organizers of previous workshops on ideation and deliberation systems at various venues such as CSCW, COOP, and C&T. These characteristics of the organizing team imply that SIG attendees will benefit from a well-balanced set of topics and a broad range of research experiences and perspectives. Specifically, Lu Xiao is an information scientist with a keen research interest on the role of the shared rationales in online deliberations, and technologies to support rationale extraction, rationale articulation and reuse. Weiyu Zhang is a communication scholar interested in exploring the connections between technological innovations and deliberative processes. Anna Przybylska is a programme manager of the Centre for Deliberation at the Institute of Sociology, University of Warsaw; and a manager of an applied project "New perspectives for dialogue: The model of deliberation and ICT tools in decision-making processes" financed by the National Centre for Research and Development within the Social Innovations Programme. Gregorio Convertino is Senior User Research at Informatica, co-editor of a recent special issue "Large-Scale Ideation and Deliberation: Tools and Studies in Organizations" in the journal of Social Media in Organizations. Also a co-editor of the special issue, Anna De Liddo is a Research Fellow in collective intelligence infrastructures and has an interest in knowledge construction through discourse, and the role of technology in scaffolding dialogue and argumentation in contested domains. Todd Davies co-edited the book *Online Deliberation: Design, Research, and Practice* (2009). His research lies in the intersections of cognition, computation, deliberation, decision-making, and informatics. Mark Klein is a Principal Research Scientist at the MIT Center for Collective Intelligence and led the development of The Deliberatorium, a platform that enables online large-scale deliberation.